\documentclass[10pt,twoside,showpacs, superscriptaddress,twocolumn]{revtex4}
\usepackage{graphicx,latexsym,amssymb,amsmath,color, multirow, mathrsfs, booktabs}

\usepackage[section]{placeins}

\newcommand{\beq}{\begin{equation}}
\newcommand{\eeq}{\end{equation}}
\newcommand{\beqn}{\begin{eqnarray}}
\newcommand{\eeqn}{\end{eqnarray}}
\newcommand{\bea}{\begin{array}}
\newcommand{\eea}{\end{array}}
\newcommand{\bsub}{\begin{subequations}}
\newcommand{\esub}{\end{subequations}}
\newcommand{\bpm}{\begin{pmatrix}}
\newcommand{\epm}{\end{pmatrix}}

\newcommand{\ff}[1]{\frac{1}{#1}}

\newcommand{\lrl}[1]{\left|#1\right|}
\newcommand{\lrb}[1]{\left(#1\right)}
\newcommand{\lrs}[1]{\left[#1\right]}

\newcommand{\Lrb}[1]{\left\{#1\right\}}

\newcommand{\svec}[1]{{\mbox{\boldmath${ #1}$}}}
\newcommand{\ivec}{\vec}

\newcommand{\figref}[1]{{\sf\bfseries Fig. \ref{#1}}}

\begin{document}

\title{Evolution of Nuclear Shell Structure due to the Pion Exchange Potential}
\author{WenHui Long}\email{whlong@pku.org.cn}
 \affiliation{School of Physics, Peking University, 100871 Beijing, China}
 \affiliation{Center for Mathematical Sciences, University of Aizu, Aizu-Wakamatsu, 965-8580
Fukushima, Japan}
\author{Hiroyuki Sagawa}
 \affiliation{Center for Mathematical Sciences, University of Aizu, Aizu-Wakamatsu, 965-8580
Fukushima, Japan}
\author{Jie Meng}
 \affiliation{School of Physics, Peking University, 100871 Beijing, China}
\author{Nguyen Van Giai}
 \affiliation{Institut de Physique Nucl$\acute{e}$aire, CNRS-IN2P3, Universit$\acute{e}$ Paris-Sud,
91406 Orsay, France}

 \begin{abstract}
The evolution of nuclear shell structure is investigated for the first time within density-dependent relativistic
Hartree-Fock theory and the role of $\pi$-exchange potential is studied in detail. The energy differences between the
neutron orbits $\Lrb{\nu1h_{9/2},\nu 1i_{13/2}}$ in the $N=82$ isotones and between the proton ones
$\Lrb{\pi1g_{7/2},\pi1h_{11/2}}$ in the $Z=50$ isotopes are extracted as a function of neutron excess $N-Z$. A kink
around $Z = 58$ for the $N=82$ isotones is found as an effect resulting from pion correlations. It is shown that the
inclusion of $\pi$-coupling plays a central role to provide realistic isospin dependence of the energy differences. In
particular, the tensor part of the $\pi$-coupling has an important effect on the characteristic isospin dependence
observed in recent experiments.
 \end{abstract}
\maketitle



The nucleon-nucleon bare interaction at low and medium energies is originally induced by the meson exchange processes
as predicted by Yukawa \cite{Yukawa:1935}. In the nuclear medium, it is strongly renormalized by medium effects which
lead to the effective nucleon-nucleon interaction. A large part of the present understanding of nuclear structure is
based on self-consistent mean field descriptions making use of effective interactions directly parametrized so as to
reproduce selected nuclear properties. The two most successful categories are the non-relativistic Hartree-Fock
approaches \cite{Vautherin:1972,Beiner:1975,Chabanat:1997,Decharge:1980} and the Relativistic Mean Field (RMF)
approaches \cite{Miller:1972, Walecka:1974, Serot:1986}. Using these models, many nuclear structure properties  are
calculated in the whole region of the nuclear chart with these effective interactions. During the past decade, great
successes have been achieved by the RMF theory not only in stable nuclei but also in exotic regions
\cite{Reinhard:1989, Ring:1996, Lalazissis:1997, Serot:1997, Typel:1999,Niksic:2002, Bender:2003,Long04, Meng:2006}. Of
special interest is the fact that the RMF model provides a natural mechanism for explaining the spin-orbit splittings
of single-particle levels. This feature becomes even more of central importance with the experimental observation that
nuclei near drip lines undergo modifications of their shell structure, where the spin-orbit potential must play an
essential role.

One of the basic open problems is the role of one-pion exchange process, which is known to play a fundamental role in
the meson-exchange interaction. However, the RMF model is not the appropriate framework to study pion-related processes
because it is essentially a Hartree approximation where  the Fock (exchange) contributions are altogether dropped,
while the Hartree (direct) contributions of pions are zero due to the parity conservation in spherical and axially
deformed nuclei. Recent progress in the relativistic Hartree-Fock description of nuclear structure, namely the
density-dependent relativistic Hartree-Fock (DDRHF) approach \cite{Long:2006} has brought a new insight to consider
this problem. Within the DDRHF theory, the effective meson-nucleon coupling strengths including the one-pion exchange
are determined in a similar way to the RMF model and a quantitative description of the nuclear structure properties can
be successfully achieved \cite{Long:2006a, Long:2006PS}. Thus, DDRHF opens the possibility to investigate the role of
one-pion exchange processes in nuclear structure problems within the framework of a relativistic mean field theory.

In this work, we study the role of pion-exchange processes on single-particle spectra by concentrating on specific
cases. In the recent paper of Schiffer et al. \cite{Schiffer:2004}, it was shown that a set of states outside the
proton core $Z = 50$ and the neutron core $N=82$ may provide a unique information to determine the evolution of the
nuclear shell structure. This is why we choose here to discuss the pion effects taking these nuclei as an illustrative
example.

The one-pion exchange process with pseudo-vector coupling contains two types of contributions, the central coupling and
the non-central tensor coupling \cite{Bouyssy:1987}. Recently, the tensor type force was shown to have a distinct
effect on the evolution of the nuclear shell structure in a non-relativistic Hartree-Fock model \cite{Otsuka:2005}. In
this work, we will study the behavior of single-particle energies of the states $\Lrb{\nu 1h_{9/2}, \nu 1i_{13/2}}$
($\nu$ denotes neutron states) in the $N=82$ isotones and the states $\Lrb{\pi 1g_{7/9}, \pi1h_{11/2}}$ ($\pi$ denotes
the proton states) in the $Z=50$ isotopes within the DDRHF theory. The effect of the $\pi$-coupling, especially the
contribution of its tensor component on the isospin dependence of the shell evolution will be analyzed in detail.

In a non-relativistic reduction, the one-pion exchange potential can
be divided into two parts, $V_\pi^c(\svec q)$ and $V_\pi^T(\svec
q)$:
 \bsub\label{Pion}\begin{align}
V_\pi^c(\svec q)= \frac{-1}3\lrs{\frac{f_\pi}{m_\pi}}^2&\frac{\svec\sigma_1\cdot\svec\sigma_2\svec q^2}{m_\pi^2+\svec
q^2}\ivec\tau_1\cdot\ivec\tau_2,\\
V_\pi^T(\svec q)= \frac{-1}3\lrs{\frac{f_\pi}{m_\pi}}^2&\frac{{3\svec\sigma_1\cdot\svec q \svec\sigma_2\cdot\svec
q-\svec\sigma_1\cdot\svec\sigma_2\svec q^2}}{m_\pi^2 + \svec q^2}\ivec\tau_1\cdot\ivec\tau_2 .
 \end{align}\esub
The Fourier-transform of the central part gives a repulsive contact
interaction and an attractive Yukawa potential,
 \beq
V_\pi^c(\svec r) =\frac{ -m_\pi^3}{12\pi}\lrs{\frac{f_\pi}{m_\pi}}^2\svec\sigma_1\cdot\svec\sigma_2
\lrs{\frac{4\pi}{m_\pi^3}\delta(\svec r) - \frac{e^{-m_\pi r}}{m_\pi r}}\ivec\tau_1\cdot\ivec\tau_2.
 \eeq
In general, the repulsive contact part is strongly hindered by the Pauli principle between two nucleons. Moreover, in
DDRHF the $\omega$-meson exchange will strongly suppress the repulsive contact part so that it can be discarded on a
solid basis \cite{Long:2006}. The central part of the $\pi$-induced potential is thus given by
 \beq\label{PionC}\begin{split}
V_\pi^c = \ff2\sum_{\alpha\beta\gamma\delta} c_\alpha^\dag  c_\beta^\dag c_\gamma c_\delta&\int d\svec r_1 d\svec r_2
\bar f_\alpha(\svec r_1) \bar f_\beta(\svec r_2)\\ \times&\lrs{\Gamma_\pi^c v_\pi}_{\lrb{1,2}} f_\gamma(\svec r_2)
f_\delta(\svec r_1),
 \end{split}\eeq
where $c_i$ and $c_i^\dag$ correspond to nucleon annihilation and creation operators, and the interaction vertex
including the propagator is
 \beq
\lrs{\Gamma_\pi^c v_\pi}_{\lrb{1,2}} = {\ivec\tau_1\cdot\ivec\tau_2}\lrs{ f_\pi \svec\gamma\gamma_5}_1\cdot\lrs{ f_\pi
\svec\gamma\gamma_5}_2 \frac{e^{-m_\pi\lrl{\svec r_1-\svec r_2}}}{12\pi\lrl{\svec r_1-\svec r_2}}.
 \eeq
The tensor $\pi$-coupling $V_\pi^T$ can be obtained similarly.

In the rest of this paper, the DDRHF calculations are performed with the parameter sets PKO1, PKO2, PKO3 given in Ref.
\cite{Long:2006a}, and they are compared with RMF (Hartree) results calculated with the parameter set PKDD
\cite{Long04}. It should be noticed that PKO3 has a stronger $\pi$-coupling than PKO1, while PKO2 and PKDD have no
$\pi$-coupling in the Lagrangian. The pairing correlations are treated by the BCS method with a density-dependent,
zero-range pairing force \cite{Dobaczewski:1996}. The nuclei studied are located along the isotones with $N=82$ from
$^{132}$Sn to $^{156}$W and the Sn isotopes with $Z=50$ from $^{104}$Sn to $^{134}$Sn.

 \begin{figure}[htbp]
\includegraphics[width = 0.40\textwidth]{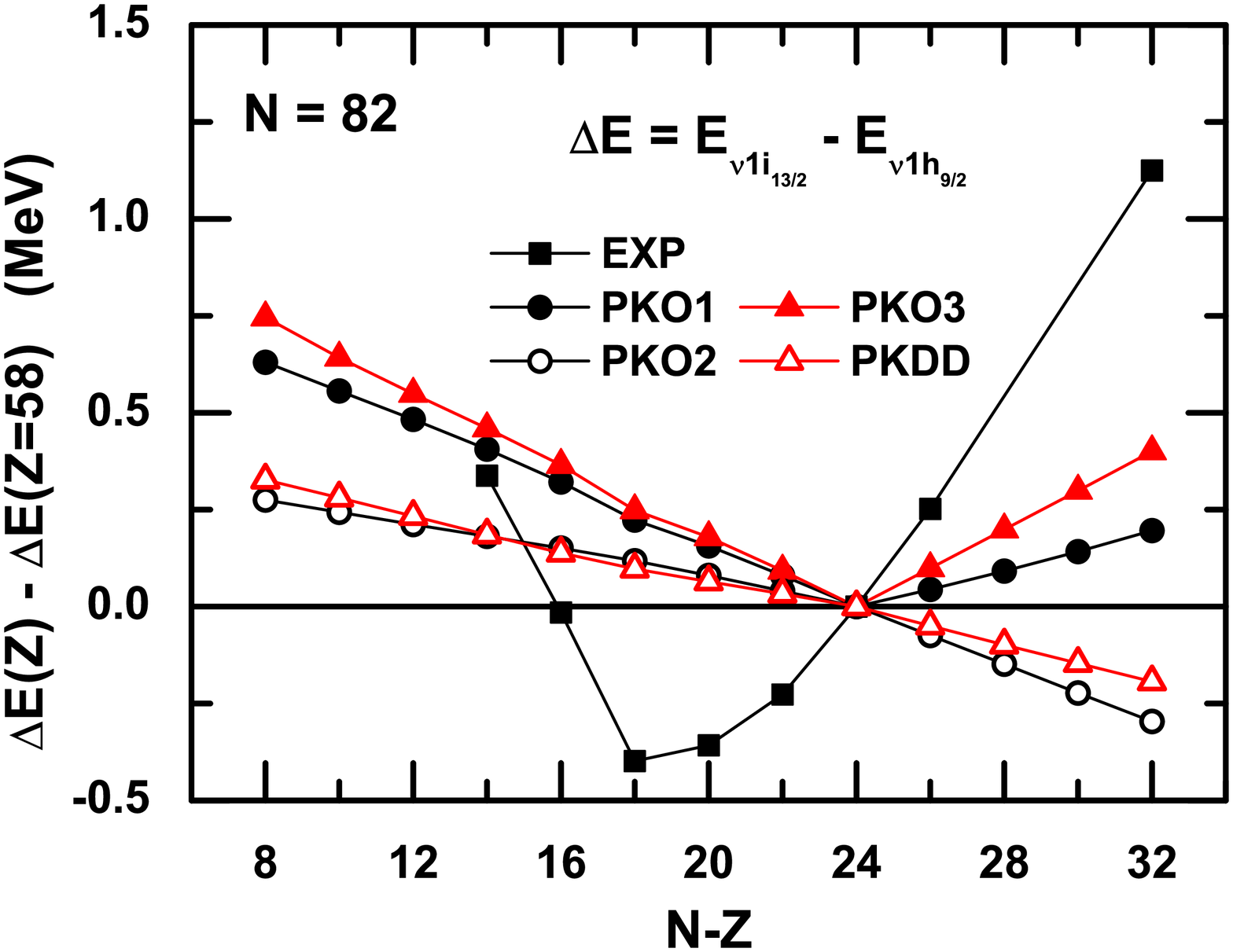}
\includegraphics[width = 0.40\textwidth]{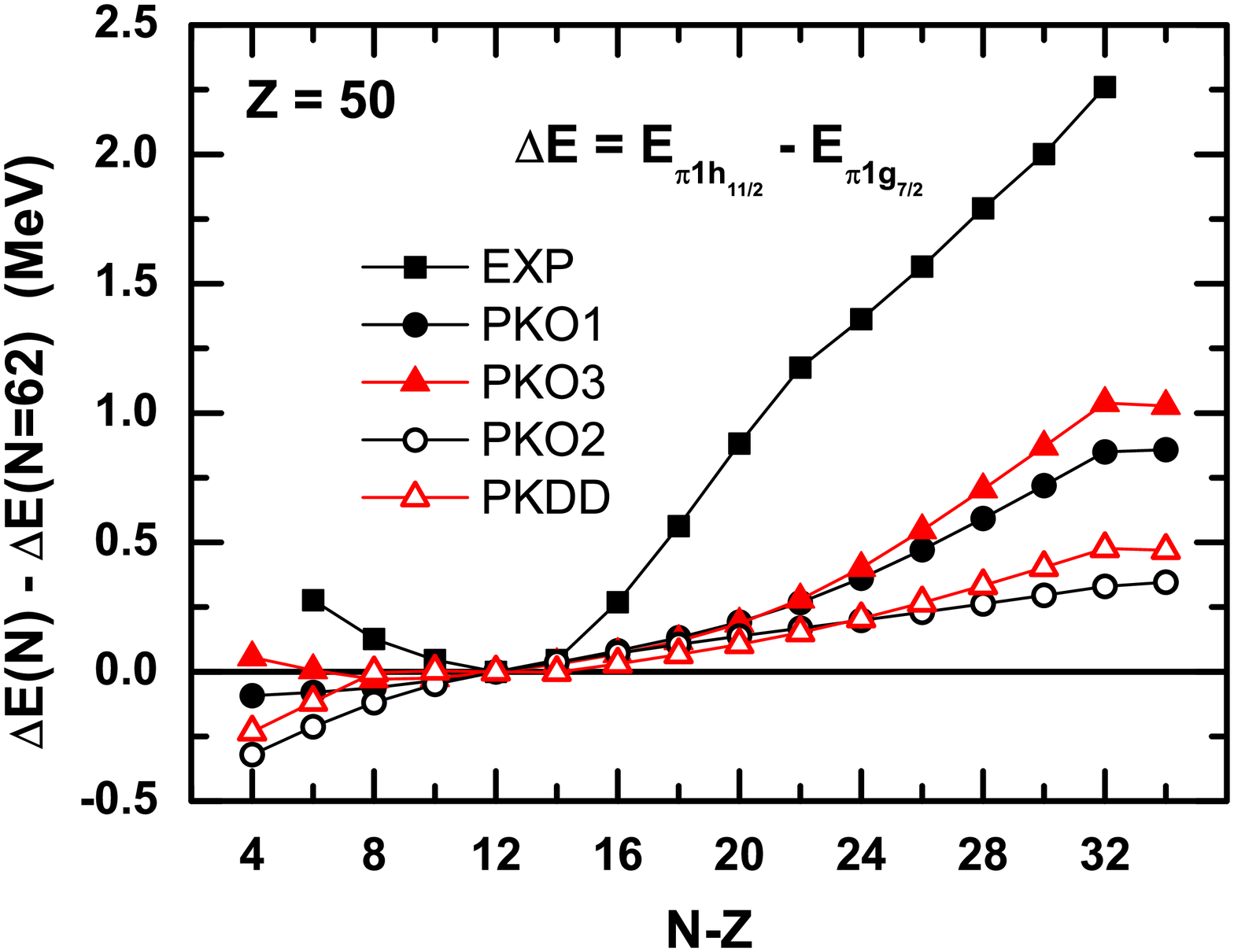}
\caption{The energy differences $\Delta E(Z)=E_{\nu 1i_{13/2}} - E_{\nu 1h_{9/2}}$ in $N=82$ isotones (first panel),
and $\Delta E(N)=E_{\pi 1h_{11/2}} - E_{\pi 1g_{7/2}}$ in $Z=50$ isotopes (second panel) as a function of neutron
excess. The DDRHF results are due to PKO1, PKO2, PKO3, while the RMF results are obtained with PKDD. The experimental
data are extracted from Ref. \cite{Schiffer:2004}.}\label{fig:ERHFD}
 \end{figure}

The energy differences $\Delta E(Z)$ between two unoccupied neutron states $\Lrb{\nu1h_{9/2},\nu1i_{13/2}}$ of the
$N=82$ isotones are shown in the first panel of \figref{fig:ERHFD} while the values $\Delta E(N)$ between two
unoccupied proton states $\Lrb{\pi1g_{7/2},\pi1h_{11/2}}$ of the $Z=50$ isotopes are shown in the second panel. The
results of pion-dependent models PKO1 and PKO3 show kinks at $Z = 58$ ($N-Z = 24$, i.e.,$^{140}$Ce) for the $N=82$
isotones. In contrast, the pion-independent models PKO2 and PKDD predict a  monotonous decrease with respect to $N-Z$.
The predictions of PKO1 and PKO3 give more realistic isospin dependence, qualitatively and quantitatively, than those
of PKO2 and PKDD as compared with the experimental data, although substantial discrepancies still exist. A similar
isospin dependence is also observed for the proton states of the Sn isotopes in the second panel of \figref{fig:ERHFD}:
the inclusion of $\pi$-coupling in RHF (PKO1 and PKO3) brings also a significant improvement compared with the results
calculated without $\pi$-coupling (PKO2 and PKDD). Around $^{112}$Sn there exists a small kink in the data and only
PKO3 gives such a trend, which might be due to its stronger $\pi$-coupling than PKO1.

\begin{figure}[htbp]
\includegraphics[width = 0.40\textwidth]{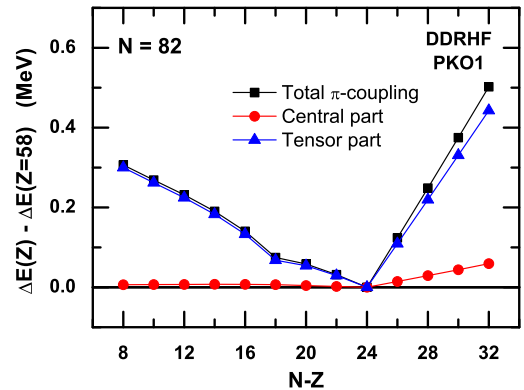}
\caption{Central and tensor pion contributions to the energy differences $\Delta E(Z) = E_{\nu1i_{13/2}} -
E_{\nu1h_{9/2}}$ due to the $\pi$-coupling, as a function of $N-Z$ for the $N=82$ isotones. The results are calculated
by DDRHF with PKO1.}\label{fig:Pion1}
 \end{figure}

The single-particle energies contain contributions from the different mesons. The pion contributions can in turn be
separated into a central and a tensor component. In \figref{fig:Pion1} are shown the $\pi$-coupling contributions to
the $\Delta E(Z)$ in the $N=82$ isotones calculated with PKO1. Comparing to the first panel of \figref{fig:ERHFD}, one
can see that the characteristic isospin dependence of $\Delta E$, namely the kink at $Z=58$, is mainly due to the pion
coupling and particularly to its tensor part.

 \begin{figure}[htbp]
\includegraphics[width = 0.40\textwidth]{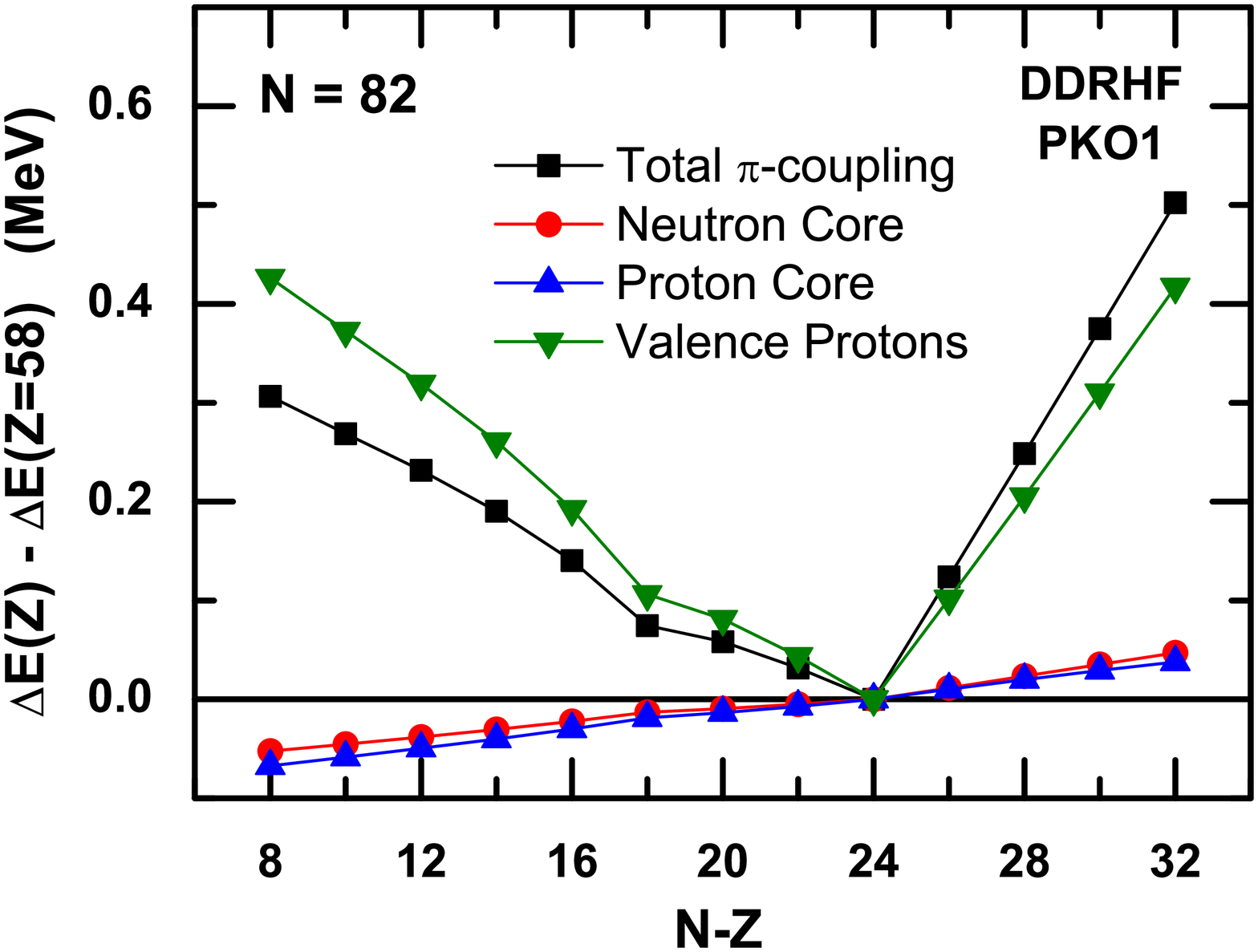}
\includegraphics[width = 0.40\textwidth]{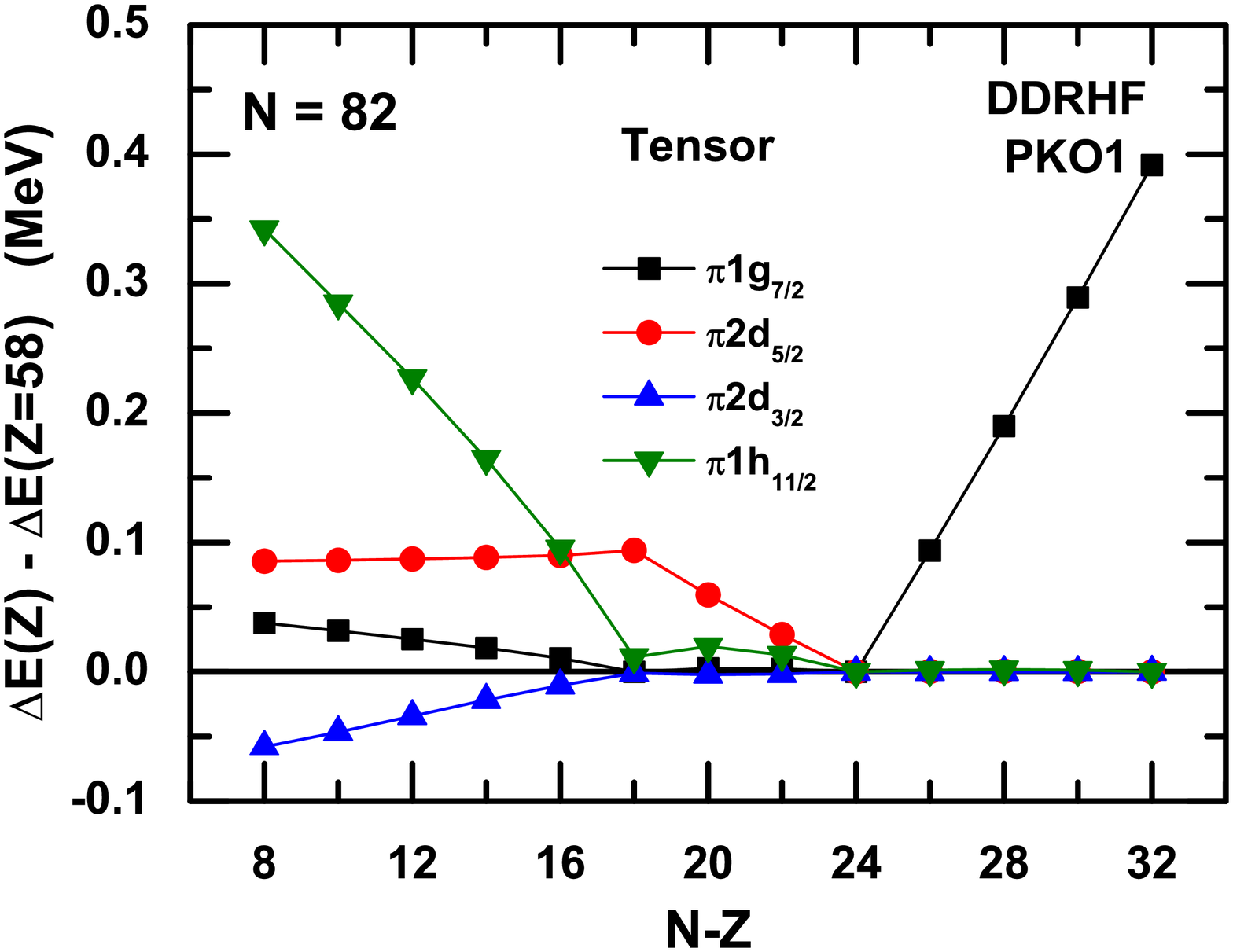}
\caption{The contribution of $\pi$-coupling to the energy difference $\Delta E = E_{\nu1i_{13/2}} - E_{\nu1h_{9/2}}$ as
a function of $N-Z$ along the isotonic chain $N=82$. In the first panel, it is separated into three parts: Neutron Core
($N=82$), Proton Core ($Z=50$) and Valence Protons. The contributions of the tensor $\pi$-coupling from 4 valence
proton states are shown in the second panel. The results correspond to DDRHF with PKO1.}\label{fig:Pion2}
 \end{figure}

We can deepen further our understanding by examining the configuration dependence of the $\pi$-coupling contributions.
From $^{132}$Sn to $^{156}$W the added protons will gradually occupy the valence orbits $\pi 1g_{7/2}, \pi 2d_{5/2},
\pi 2d_{3/2}$ and $\pi1h_{11/2}$. In \figref{fig:Pion2} we separate the contributions of $\pi$-coupling to $\Delta
E(Z)$ into three components from the neutron core orbits ($N=82$), the proton core orbits ($Z=50$) and the valence
proton orbits, respectively. From the first panel of \figref{fig:Pion2}, it is seen that the valence protons play the
most important role to give the isospin dependence shown in \figref{fig:ERHFD} and \figref{fig:Pion1}. In the second
panel of \figref{fig:Pion2}, the contributions of 4 valence proton orbits to $\Delta E(Z)$ are separately plotted to
examine the configuration dependence of the tensor $\pi$-coupling. As the proton number increases from $Z=50$ to 58
($N-Z= 32$  to 24), the $\pi1g_{7/2}$ orbit is gradually occupied. Then, from $Z=60$ to 64 ($N-Z= 22$  to 18), the
protons fill the $\pi2d_{5/2}$ orbit and finally the $\pi1h_{11/2}$ orbit is occupied in the nuclei with $Z=66\sim 74$
($N-Z = 16$ to 8). It is found from the second panel of \figref{fig:Pion2} that the magnitude of the tensor correlation
is very sensitive to the occupation probabilities of the valence orbits, especially, those of high-$j$ orbits.

\begin{figure}[htbp]
\includegraphics[width = 0.40\textwidth]{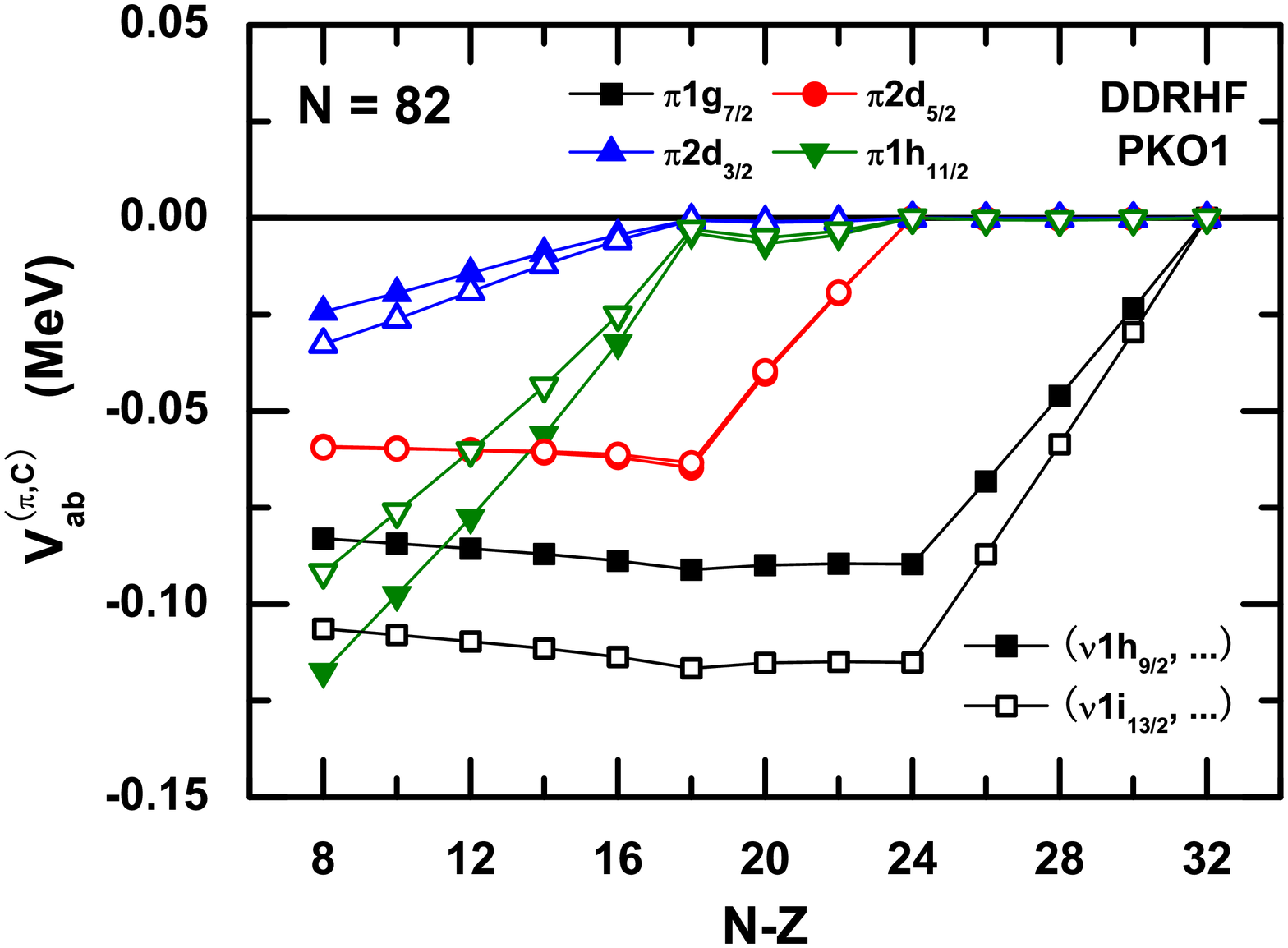}
\includegraphics[width = 0.40\textwidth]{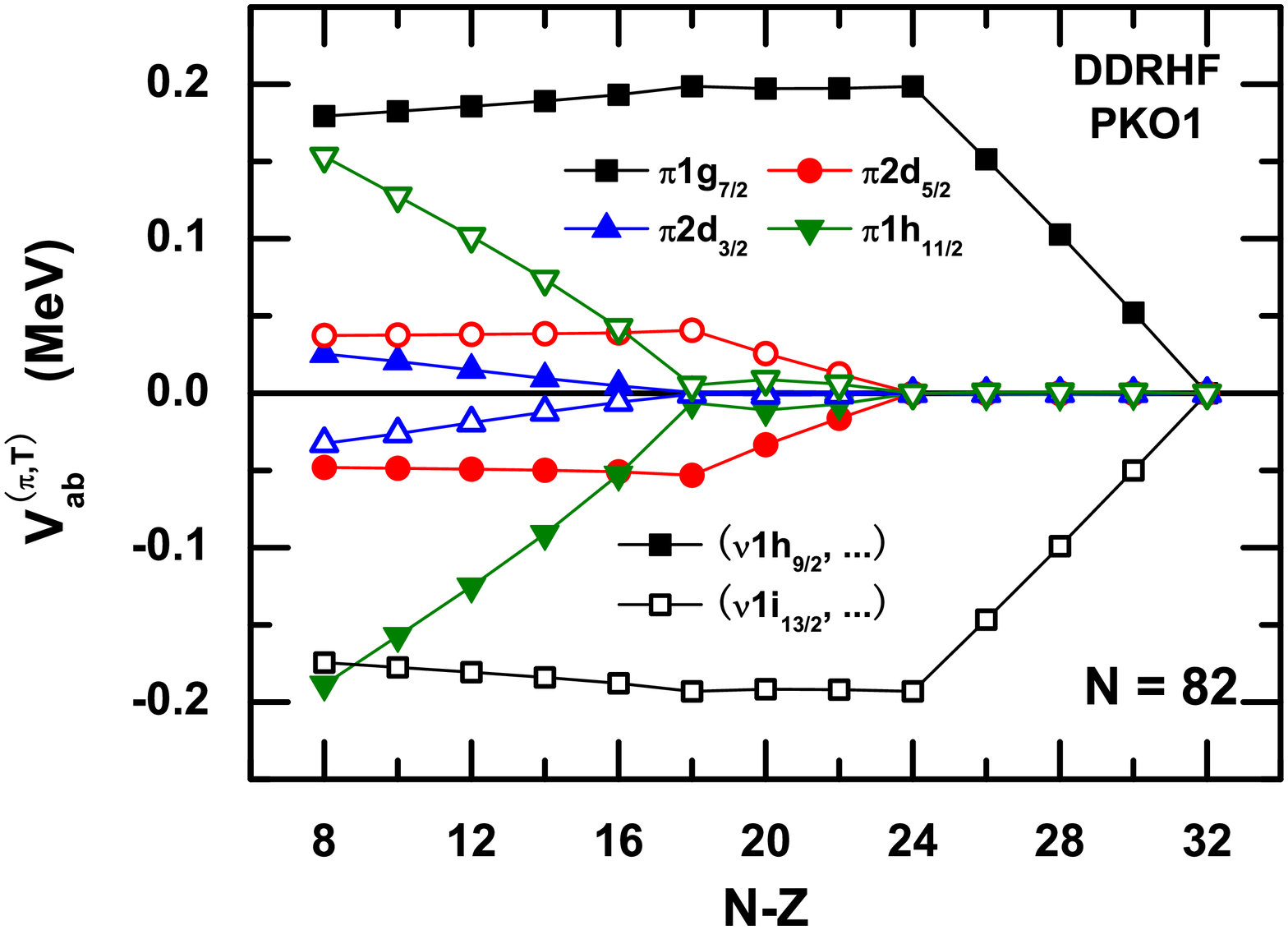}
\caption{The central (first panel) and tensor (second panel) parts of the interaction matrix elements between neutron
states ($\nu 1h_{9/2}$ (filled symbols) and $\nu1i_{13/2}$ (open symbols)) and valence proton states ($\pi 1g_{7/2}$
(squares), $\pi 2d_{5/2}$ (circles), $\pi 2d_{3/2}$ (up-triangles) and $\pi1h_{11/2}$ (down-triangles)) induced by the
$\pi$-coupling as function of $N-Z$ in the $N=82$ isotones. The results correspond to DDRHF with
PKO1.}\label{fig:VpiCT}
 \end{figure}

To have a better understanding about the contributions of these valence proton states, we calculate the interaction
matrix elements of $\pi$-coupling, separating the central and tensor parts, $V_{ab}^{(\pi,\;c)}$ and
$V_{ab}^{(\pi,\;T)}$. \figref{fig:VpiCT} shows the central (first panel) and tensor (second panel) matrix elements
between the neutron states, $\nu1h_{9/2}$ or $\nu1i_{13/2}$ and the different valence proton states. The central
$\pi$-coupling is attractive and it shows a simple monotonous isospin dependence for both $\nu1h_{9/2}$ and
$\nu1i_{13/2}$. In contrast, the tensor part shows different isospin behaviours depending on the configurations.
Namely, the two-body matrix elements between $\nu1h_{9/2}$ and $\pi1g_{7/2}$ or $\pi2d_{3/2}$ are repulsive whereas
those between $\nu1h_{9/2}$ and $\pi1h_{11/2}$ or $\pi2d_{5/2}$ are attractive. For $\nu1i_{13/2}$, they have
completely opposite signs as seen in the second panel of \figref{fig:VpiCT}.

The first panel of \figref{fig:VpiCT} shows that the central $\pi$-coupling is always attractive. As a result, it
provides the simple monotonous behaviour and weak isospin dependence (see \figref{fig:Pion1}). At first sight, the
tensor $\pi$-coupling looks more complicated. However, one can find the following regularities: 1) the isospin
dependence of the matrix elements is determined by the occupation probabilities of the valence proton orbits. 2) the
tensor $\pi$-coupling is attractive for $\lrb{j_<, j_>}$ configurations, while it is repulsive for $\lrb{j_>, j_>}$ and
$\lrb{j_<, j_<}$ configurations. 3) the magnitude of the tensor interaction matrix element is proportional to the
degeneracy $(2j+1)$ of the orbit. These properties of the tensor force in DDRHF with PKO1 and PKO3 can provide a more
reasonable isospin dependence of the energy difference between $\nu1h_{9/2}$ and $\nu1i_{13/2}$. This peculiar feature
of the tensor interaction was already pointed out in the deuteron binding mechanism in Ref. \cite{Blatt:1952}, and also
in the non-relativistic HF calculations \cite{Otsuka:2005, Dobaczewski:1996}.

In this work, the evolutions of nuclear shell structure outside the $N=82$ neutron and $Z=50$ proton closed shells are
investigated for the first time within the density-dependent relativistic Hartree-Fock (DDRHF) theory and the role of
the one pion-exchange potential is studied in detail. The energy differences $\Delta E(N-Z)$ between the states
$\Lrb{\nu1h_{9/2}, \nu 1i_{13/2}}$ in the $N=82$ isotones, and those between $\Lrb{\pi1g_{7/2}, \pi1h_{11/2}}$ in the
$Z=50$ isotopes are extracted as a function of neutron excess to explore the shell evolution of these exotic nuclei. It
is found that a better isospin dependence of $\Delta E(N-Z)$ is obtained if one includes a pion-coupling term in the
effective Lagrangian. This pion coupling gives the characteristic isospin dependence observed in recent experiments.
Furthermore, the tensor $\pi$-coupling with the valence protons provides the important physical mechanism to give a
proper description of the isospin dependence. Indeed, the strong configuration dependence of the tensor part of the
$\pi$-coupling plays a crucial role to give the kink of the $\Delta E(N-Z)$ observed in the experimental systematics of
the shell evolution.

Compared with the experimental results, however, there still exist some substantial discrepancies. Since the studied
states are unoccupied orbitals, the dynamical coupling to the core vibrations which is not considered here, might be
one missing physical mechanism to cure the existing disagreement. In the present DDRHF the $\pi$-nucleon effective
coupling, determined by fitting the empirical data of the ground state properties of selected nuclei, is
density-dependent and its strength turns out to be fairly weak in the medium \cite{Long:2006a}. Another promising meson
coupling for the shell evolution is the Lorentz $\rho$-tensor coupling which is not included yet. As a perspective, the
inclusion of dynamical coupling to RPA vibrations and the treatment of $\rho$-tensor coupling might play an additional
important role to cure the existing discrepancy.

\begin{acknowledgements}
{This work is  partly supported by the National Natural Science Foundation of China under Grant No. 10435010, and
10221003, and the Japanese Ministry of Education, Culture, Sports, Science and Technology by Grant-in-Aid for
Scientific Research under the program number (C(2)) 16540259, and the European Community project Asia-Europe Link in
Nuclear Physics and Astrophysics CN/Asia-Link 008(94791). }
\end{acknowledgements}

%

\end{document}